# Beam hardening correction of Bremsstrahlung X-ray computed tomography measurements and required measurement accuracies

Nikhil Deshmukh[a]

[a]*Pacific Northwest National Laboratory, Richland, Washington*

*E-mail: Nikhil.deshmukh@pnnl.gov*

# Beam hardening correction of Bremsstrahlung X-ray computed tomography measurements and required measurement accuracies

*Abstract:* Doing quantitative computed tomography (CT) using Bremsstrahlung sources requires an estimate of the spectrum emitted by the X-ray source. One method of beam hardening correction (BHC), as described by Lifton[1], first uses transmission measurements of a known material and range of thicknesses to estimate the spectrum and then uses the estimated spectrum to correct the beam hardening error, not necessarily on the same material as used for estimating the spectrum. We adopt Lifton's BHC procedure and through simulations show how the uncertainties in the transmission measurements and the uncertainties in the material thickness measurements propagate to uncertainties in the spectrum, the beam hardening correction curve and thence to uncertainties in the beam hardening corrected CT reconstruction.

## Introduction

Industrial computed tomography (CT) scanners use Bremsstrahlung X-ray sources which produce a poly-energetic spectrum of X-ray photons (unlike synchrotron sources which produce a monoenergetic X-ray beam). Poly-energetic spectra create a specific kind of effect in the CT reconstruction known as the beam hardening effect. To do quantitative work, it is essential to correct for the beam hardening effect which in turn requires having a good estimate of the spectrum. A well-known way of estimating the Bremsstrahlung spectrum is to do a series of transmission measurements with suitably selected material(s) of precisely known thicknesses and density. Unfortunately, estimating the Bremsstrahlung X-ray spectrum from a series of transmission measurements is a hard linear inverse problem because the condition numbers involved are very high, and as a result small errors in the measurements result in greatly amplified noise in the estimated spectrum. For this study, we will be choosing the same materials and thicknesses as Lifton [1] does in his paper to demonstrate beam hardening correction. As a prelude to this work, we recreated the attenuation curves and beam hardening curves as shown in that work. While it describes a very useful procedure, the Lifton paper does not quantify the accuracy required on the transmission and thickness measurements to get to a certain degree of accuracy in the resulting reconstruction after applying a beam hardening correction. The aim of this study is to use simulations to estimate how much error in the transmission measurements and in the thickness measurements can be tolerated when expecting a certain degree of accuracy in the reconstruction after BHC. However, we do deviate from the Lifton procedure in a couple of ways: we apply the beam hardening correction to an acrylic object instead of aluminum, and we use least squares solvers (the Lifton paper does not provide sufficient details on the solver they used).

## Procedure

The measurement scheme as described in the Lifton paper uses transmission measurements from titanium sheets of twenty different thicknesses varying from 0.1 mm to 6 mm. We generate a forward system matrix corresponding to these transmission measurements with 130 bins per row (1 keV bin width) and attenuation coefficients (including coherent scatter) obtained from the xraydb [2] database. Each row of the matrix corresponds to a different thickness of titanium. We also adopt the spectrum used in the Lifton paper as our nominal spectrum. A plot of the rows of the forward system matrix and the nominal spectrum are shown in Fig 1.

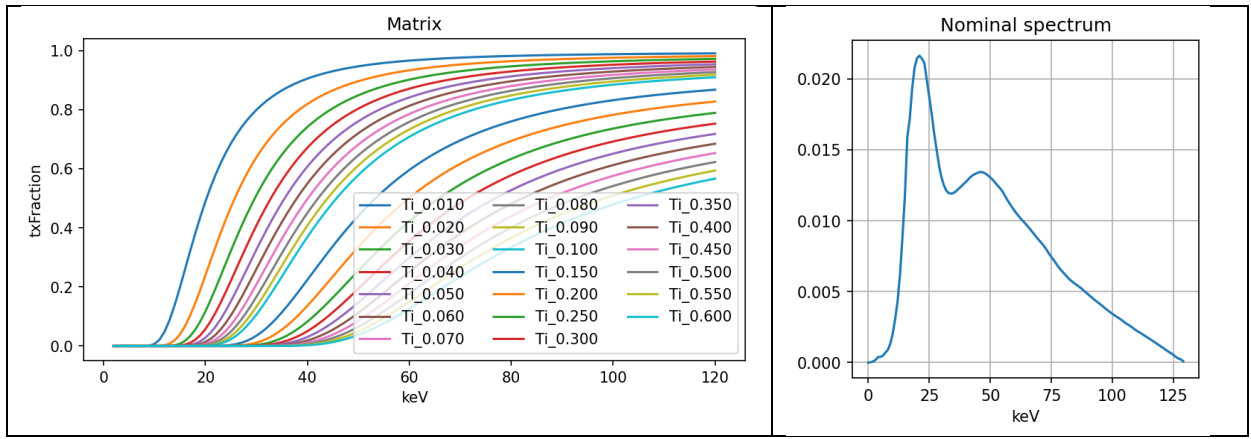

Fig 1: a) Line plots of the rows of the forward system matrix (with each row being one of the 20 titanium measurements as in the Lifton paper). b) The nominal spectrum used in the simulations.

The forward problem is written as Ax=b, where A is the matrix representing the forward model, x is the spectrum, and b is the transmission measurement vector.

We estimate the emitted spectrum by solving the inverse problem: $x = A^{-1} * b$. Errors in x come from either errors in A (which come from errors in thickness measurements) or errors in b, the transmission measurements. (There can be other sources like error in density of material(s) or error in the attenuation coefficients, which are not considered here.)

The steps to determine requirements on transmission measurements to achieve a certain accuracy in the beam hardening-corrected CT reconstruction are:

1. Generate the forward system matrix for attenuation through the titanium sheets at discrete energies as described above.
2. Generate the nominal measurement vector using the system matrix and the reference spectrum.
3. Add Gaussian noise with zero mean and known standard deviation to the nominal measurement vector to obtain a noisy measurement vector.
4. Calculate the estimated spectrum using the system matrix and noisy measurement vector.
5. Calculate the beam hardening correction curve using the estimated spectrum.
6. Correct the projection(s) using the beam hardening correction curve and do CT reconstruction.
7. Repeat steps 3, 4, 5, and 6 fifty times.

The steps to determine accuracy requirements on material thickness measurements to achieve a certain accuracy in the beam hardening-corrected CT reconstruction are:

1. Generate the nominal forward system matrix as described above using the nominal thicknesses of titanium sheets.
2. Generate the nominal measurement vector using the nominal system matrix.
3. Add Gaussian noise with zero mean and known standard deviation to the nominal thicknesses and generate a set of noisy thickness numbers.
4. Generate a noisy system matrix using the noisy thickness numbers.
5. Generate the estimated spectrum using the noisy system matrix and nominal measurement vector from step 2.
6. Calculate the beam hardening correction curve using the estimated spectrum.

7. Correct the projection(s) using the beam hardening correction curve and do CT reconstruction.
8. Repeat steps 3, 4, 5, 6, and 7 fifty times.

Details of the steps and the results are given in the following sections.

### Unfolding the spectrum

Since the Lifton paper does not provide enough details on the reconstruction algorithm used, we have chosen to do the reconstruction with an easy to implement and replicate solver. We use the Truncated Singular Value Decomposition (TSVD) solver with five singular values to do the inversions. Five singular values provided the best results for our data set. We also did the inversions using a Tikhonov regularized least squares solver and found the results to be quite similar.

### Procedure for accuracy requirements on transmission measurements

The matrix that represents the forward model of the measurement scheme is used to generate a vector of nominal simulated measurements of transmission fraction (fraction of X-rays transmitted by a sheet, varies between 0 and 1 and is unitless). Next, Gaussian noise with known standard deviation (0.8E-3) is added to each of the 20 measurements in the vector. This gives us one instance or set of noisy measurements. This is used with the TSVD solver and the matrix to generate one estimate of the X-ray spectrum. We repeat this procedure fifty times to generate fifty estimated spectra and then find the mean and standard deviation of each bin of the spectra. The level of the added noise is chosen such that the errors in the estimated spectra from the procedure have a maximum SD of ~5%. A range of transmission noise values were tried to find this noise level in the resulting spectra.

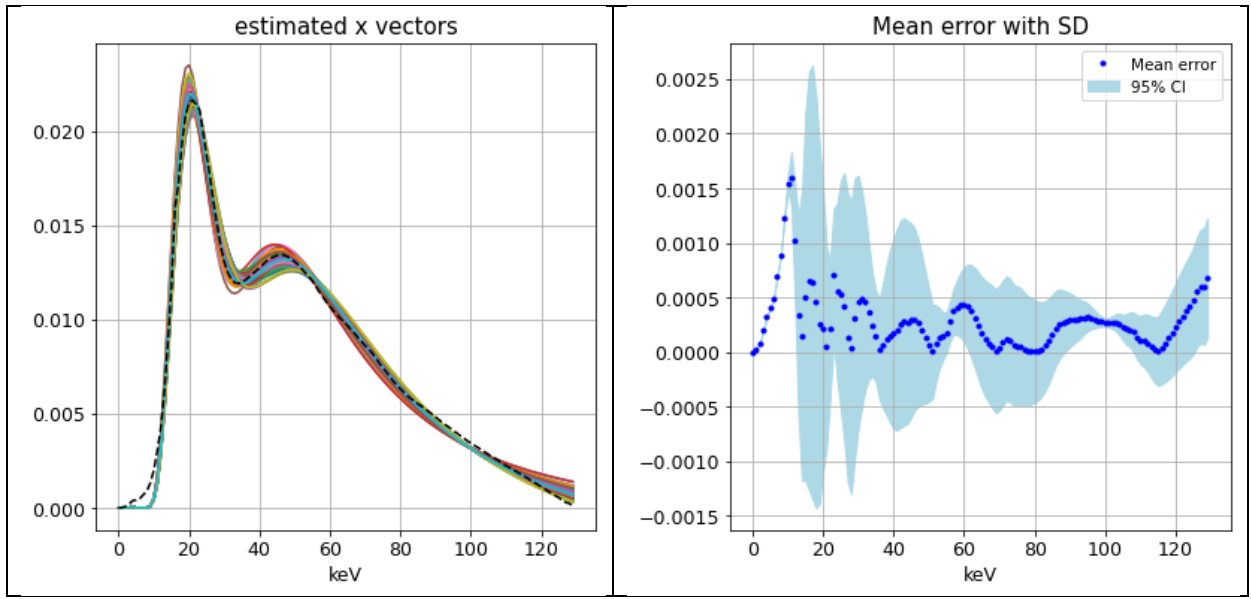

Fig 2: a) Ensemble of spectra from inversion process. The black dashed line is the reference/nominal spectrum. b) Mean and +/- two standard deviations (2*SD) of the bin-wise error in the ensemble of estimated spectra. The maximum SD is ~5% of the true spectrum. Similar ensembles are obtained from transmission measurement perturbations and thickness perturbations.

### Procedure for accuracy requirements on material thickness measurements

The matrix that represents the forward model of the measurement scheme is used to generate a vector of nominal simulated measurements. Next, Gaussian noise with known standard deviation (+/- 1 micron) is added to each of the 20 material thickness measurements used to generate a perturbed system matrix. This gives us one instance of a noisy matrix. This is used with the nominal measurement vector and the TSVD solver to generate one estimate of the X-ray spectrum. We repeat this procedure fifty times to generate fifty estimated spectra and then find the mean and standard deviation of each bin of the fifty spectra (Fig 2). The level of the added noise is chosen such that the errors in the estimated spectra from the procedure have a maximum SD of ~5%. A range of thickness noise values were tried to find this noise level in the resulting spectra.

### Using perturbed spectra to generate perturbed BHC curves

The perturbed spectra from either of the procedures above (noisy measurements [b] or noisy matrix [A]) are used along with the X-ray attenuation coefficients of acrylic to generate the beam hardening correction curves for an object made of acrylic material. Figure 3 shows the beam hardening correction from a poly-energetic spectrum to a monoenergetic spectrum; the X-axis is the attenuation due to the polychromatic X-ray spectrum as obtained from the previous step and the Y-axis is the attenuation if the beam had been monochromatic. See Lifton paper for details about how the monoenergetic energy value is chosen.

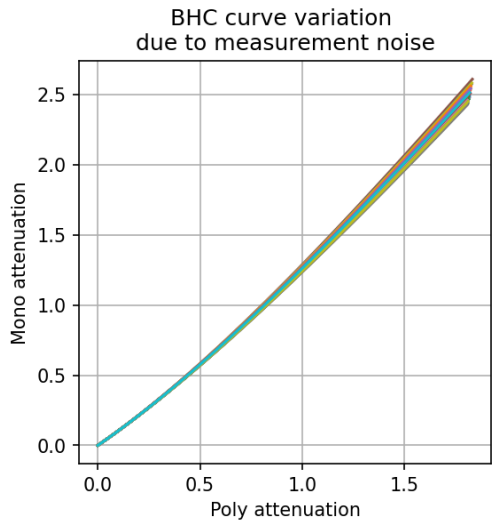

Fig 3: Ensemble of Beam Hardening Correction curves for acrylic obtained from ensemble of spectra resulting from solving the inverse problem. Similar ensembles are obtained from transmission measurement perturbations and thickness perturbations.

### Using perturbed spectra to generate ensemble of corrected CT reconstructions

Simulated projection(s) for an acrylic cylinder (3 inches tall, 2 inches diameter) are obtained from an MCNP simulation done using CardSharp [3], see Fig 4a. The projection(s) come from an MCNP6 FIR5 tally with 300 x 300 pixels. If a CT reconstruction is made from these projections without any correction for beam hardening, a severe cupping artifact is seen, as shown in Fig. 4b.

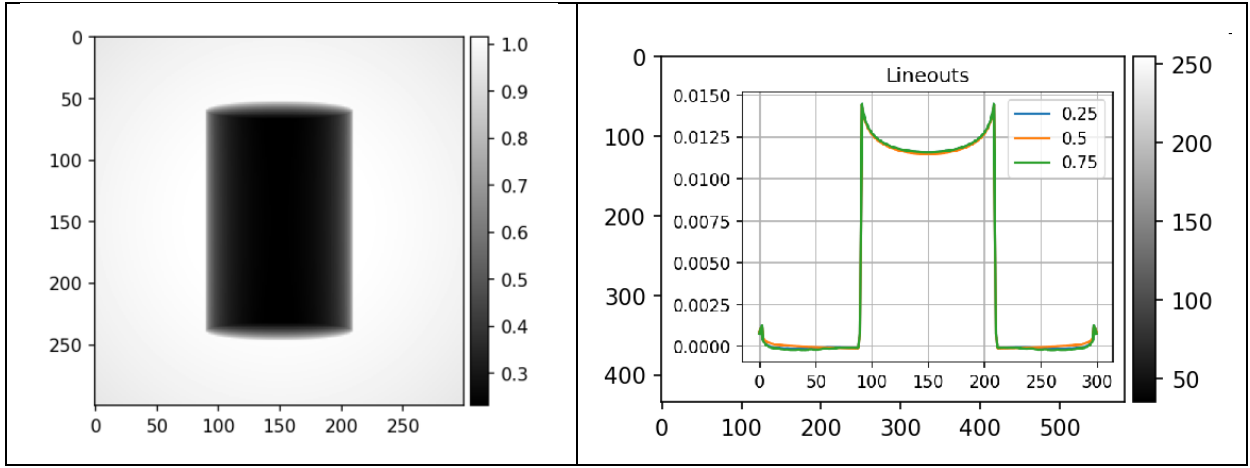

Fig 4: a) Projection of acrylic cylinder obtained from MCNP simulation with nominal spectrum. b) Line plots through the middle of the CT reconstruction of acrylic cylinder without beam hardening correction at three different heights. Note the severe cupping artifact caused by beam hardening.

The beam hardening correction curves described in the previous section are used to correct the CT projection(s) of the acrylic cylinder, and then the cylinder is reconstructed using the FDK algorithm for cone beam CT. Fifty such reconstructions are made corresponding to the fifty beam hardening correction curves from the previous step. For each reconstruction, a line plot from the central slice is plotted to see how well the correction worked (Fig 5).

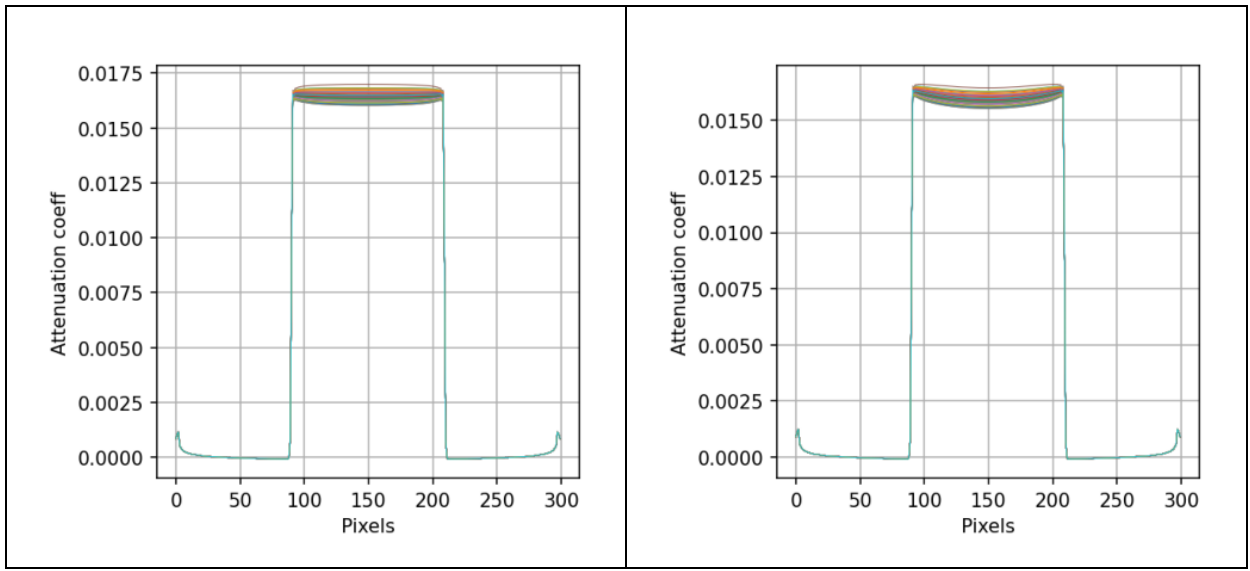

Fig 5: a) Line plots through the middle of the CT reconstruction of the acrylic cylinder with beam hardening correction (projections without scatter). b) Line plots of reconstruction with beam hardening correction (projections with scatter). Similar ensembles are obtained from transmission measurement perturbations and thickness perturbations.

When there is no scatter in the CT projection, the plots of the beam hardening corrected reconstructions have an error with a SD of ~2.5% of the true value around a mean value of zero. The amount of variation in error is found to be proportional to the amount of variation in the ensemble of the beam hardening correction curves.

When there is scatter in the CT projection, the plots of the beam hardening corrected reconstructions have some uncorrected error since the beam hardening corrected curves are not designed to correct for scatter. The amount of variation in error still depends on the amount of variation in the ensemble of the beam hardening corrected curves, but the non-zero mean error comes from the scatter in the projection(s) (scatter causes the attenuation coefficient in the interior of the object to be underestimated, just like beam hardening).

Results and discussion

We have shown that for the Beam Hardening corrected CT reconstruction of an acrylic cylinder to be within +/- 2.5% requires the error (SD) in the spectrum to be less than +/- 5%. This indirectly requires an uncertainty better than +/- 0.8E-3 for the transmission fraction measurements used for spectrum estimation and better than +/- 1 micron uncertainty in the thickness measurements of titanium sheets used for spectrum estimation. This assumes that the CT projections are free of scatter.

Since our simulations assumed only one source of error at a time, the measurement accuracy requirements when multiple sources of error are active at the same time, as would happen in practice, might be higher. We plan to investigate combinations of multiple errors sources in future work to address this. Additionally, investigations with more sophisticated solvers incorporating better a-priori estimates of the spectrum are also planned since least squares solvers clearly impose a very high requirement on the accuracy of the measurements.